\documentclass[twocolumn,showpacs,preprintnumbers,amsmath,amssymb]{revtex4}

\usepackage{graphicx}
\usepackage{dcolumn}
\usepackage{bm}

\newcommand{\pd}{\partial}

\newcommand{\nn}{\nonumber}
\newcommand{\e}{{\rm e}}

\newcommand{\del}{\delta}
\newcommand{\ra}{\rangle}
\newcommand{\la}{\langle}
\newcommand{\rar}{\rightarrow}

\newcommand{\pdm}{\partial_{\mu}}
\newcommand{\pdn}{\partial_{\nu}}
\newcommand{\fmn}{F_{\mu\nu}}
\newcommand{\am}{A_{\mu}}
\newcommand{\an}{A_{\nu}}

\newcommand{\db}{\delta_{\rm B}}
\newcommand{\bdb}{\bar{\delta}_{\rm B}}

\newcommand{\fr}{\frac}

\newcommand{\Del}{\Delta}

\begin{document}

\preprint{KUCP-0179}
\preprint{hep-th/0103075}

\title{
Phase structure of a compact $\boldsymbol{U(1)}$ gauge theory \\
from the viewpoint of a sine-Gordon model
}
\author{Kentaroh Yoshida and Wataru Souma}
 \altaffiliation[Also at ]{
Human Information Science Laboratories,
ATR International, Kyoto, 619-0288, Japan
}
\affiliation{%
Graduate School of Human and Environmental Studies,\\ 
Kyoto University, Kyoto 606-8501, Japan. \\
{\tt E-mail: yoshida@phys.h.kyoto-u.ac.jp \\ 
souma@atr.co.jp}}%

\date{\today}

\begin{abstract}
We discuss the phase structure of the four-dimensional 
compact $U(1)$ gauge theory at
finite temperature using a deformation of the topological model. 
Its phase structure 
can be determined by the behavior of the Coulomb gas (CG) 
system on the cylinder. 
 We utilize the relation between the CG system 
and the sine-Gordon (SG) model, and
 investigate the phase structure of the gauge theory in terms of the SG
 model. Especially, the critical-line equation of the gauge theory
 in the strong-coupling
 and high-temperature region is obtained by calculating the one-loop 
effective
 potential of the SG model.
\end{abstract}

\maketitle

\section{Introduction}

Recently the scenario of treating a gauge theory as a deformation of
 the topological model 
has been proposed by several authors \cite{kondo,HT,izawa}. 
The motivation of this scenario is to investigate the confinement and
the phase structure at zero temperature or finite temperature. 
In particular, we can calculate the expectation
value of the Wilson loop (at zero temperature) or
Polyakov loop (at finite temperature) by considering 
the topological model and so derive the linear
potential, which means the quark confinement \cite{wilson}. 
The Parisi-Sourlas (PS)
 dimensional reduction \cite{PS} is very powerful to study the topological
 model. This
scenario can be also applied to a compact $U(1)$ gauge theory.
It is quite related to QCD by the use of the Abelian projection, 
which is a partially gauge fixing method \cite{tHooft,kondo2}.
In the case of the compact $U(1)$ gauge theory 
the topological model becomes the two-dimensional $O(2)$ nonlinear sigma 
model (NLSM$_2$) and we can show that the confining phase exists
 in the strong-coupling
region at zero temperature and finite temperature \cite{qed,KY2}.

In the case of zero temperature, the confining-deconfining
phase transition of the compact $U(1)$ gauge theory can be described by
the Berezinskii-Kosterlitz-Thouless
(BKT) phase transition \cite{BKT} in the $O(2)$ NLSM$_2$ \cite{qed}. 
It is well known that the $O(2)$ NLSM$_2$ has vortex solutions
and is equivalent to 
several models, such as the Coulomb gas (CG) system, 
sine-Gordon (SG) model and massive Thirring (MT) model. 
In the compact $U(1)$ gauge theory 
the confining phase exists at the strong-coupling region due to the
effect of the vortex solution, which induces the linear potential
between the static charged test particles. The confining phase
transition corresponds to the BKT phase transition in the CG
system \cite{BKT}, or the Coleman transition in the SG
model \cite{Coleman}. The CG system has a phase transition from a dipole
phase to
a plasma phase.  
The quantum SG model undergoes a phase transition from a stable
vacuum to an unstable vacuum at the certain critical coupling.
 Both phase transitions are intimately connected through 
the equivalence between the CG system and the SG model.

In our previous papers \cite{KY2,KY}, we have investigated the 
phase structure of the 
compact $U(1)$ gauge theory at finite temperature from the viewpoint of the
behavior of the CG system on the cylinder. In particular, we could study
its strong-coupling and high-temperature region by using the behavior of
the one-dimensional
CG system \cite{1C}. This result is consistent with 
 the prediction in Ref. \cite{SV}. 

In this paper we would like to investigate the phase structure of the
compact $U(1)$ gauge theory at finite temperature more quantitatively from
the different aspect. 
We investigate its
phase structure from the viewpoint of the SG model by the use of
relationship the CG
system and the SG model. 
The one-loop effective potential of the SG model enables us to investigate the
phase structure of the gauge theory at the high-temperature and 
strong-coupling region. Especially, we can evaluate 
the critical-line equation. 

Our paper is organized as follows. Section II is devoted to a review of  
a deformation of a topological model. In Sec. III we discuss the equivalence between
 the thermal SG model and the CG system on the cylinder.
In Sec. IV 
the one-loop effective potential of the SG model is discussed. Especially, 
the critical-line equation of the SG model can be  obtained. 
 We can evaluate the critical-line equation of the compact $U(1)$ gauge
theory at finite temperature from this result. 
Section V  is devoted to the conclusion and discussion.

\section{Compact $\boldsymbol{U(1)}$ gauge theory as a deformation of a topological model}

In this section we introduce the method of the decomposition of the
compact $U(1)$ gauge theory  
into the perturbative deformation part and the topological model part 
[topological quantum field theory (TQFT) sector]. The 
perturbative deformation part is topologically trivial but the TQFT
sector is nontrivial. The TQFT sector has the information of the
topological objects such as vortices and monopoles, which are assumed to
play an important role in the confinement or phase transition. 
The dynamics of the confinement is encoded in the TQFT sector. 
 Therefore we can derive the linear potential by analyzing the TQFT
sector through the PS dimensional reduction \cite{PS}, which reduces the
four-dimensional TQFT sector to the two-dimensional $O(2)$
NLSM$_2$. 
If we consider the finite-temperature system then the two-dimensional 
space on which the reduced theory lives is the cylinder.

\subsection{Setup}

The action of the (compact) $U(1)$ gauge theory on the 
(3+1)-dimensional Minkowski
space-time is given by 
\begin{eqnarray}
 S_{\rm U(1)} &=& - \fr{1}{4}\int d^4x \fmn[A] F^{\mu\nu}[A], \\
 \fmn[A] &=& \pdm\an - \pdn\am.
\end{eqnarray}
The partition function is given by 
\begin{eqnarray}
\label{part}
 Z_{\rm U(1)} &=& \int [d\am][dC][d\bar{C}][dB]\exp\left(iS_{\rm U(1)} + iS_{\rm J}\right),  \\
 S_{\rm J} &=& \int d^4x \left(J^{\mu}\am + J_{\rm c}C + J_{\rm \bar{C}}\bar{C} + J_{\rm B}B\right).
\end{eqnarray}
Here we use the Becchi-Rouet-Stora-Tyutin (BRST) quantization. 
Incorporating the
(anti-) Faddeev-Popov (FP) ghost field $C (\bar{C})$ and 
the auxiliary field $B$, we can
construct the BRST transformation $\db$,
\begin{eqnarray}
 & & \db \am = \pdm C,~~~~\db C = 0, \nn \\
 & & \db \bar{C} = iB,~~~~\db B =0.  
\end{eqnarray}
The gauge fixing term can be constructed from the BRST transformation
$\db$ as 
\begin{eqnarray}
 S_{\rm GF+FP} = -i\db \int d^4x G_{\rm GF + FP}[\am,C,\bar{C},B],
\end{eqnarray}
and $G_{\rm GF + FP}$ is chosen as 
\begin{eqnarray}
\label{2.8}
 G_{\rm GF + FP} = \bdb \left[\fr{1}{2}\am A^{\mu} + iC \bar{C}\right],
\end{eqnarray}
where  $\bdb$ is the anti-BRST transformation, which is defined by 
\begin{eqnarray}
& & \bdb \am = \pdm \bar{C},~~~~\bdb C = i\bar{B}, \nn \\
& & \bdb \bar{C} = 0,~~~~\bdb \bar{B} = 0,~~~~B + \bar{B} = 0. 
\end{eqnarray}
The above gauge fixing condition (\ref{2.8}) is 
 convenient to investigate the TQFT
sector. 

We decompose the gauge field as 
\begin{eqnarray}
\label{dec}
 \am (x) &=& V_{\mu}(x) + \Omega_{\mu}(x)~(\equiv V_{\mu}^{U}),\nn\\
\Omega_{\mu} (x)& \equiv& \fr{i}{g}U(x)\pdm U^{\dagger}(x),
\end{eqnarray} 
where the $g$ is the gauge coupling constant.
Using the Faddeev-Popov determinant $\Delta_{\rm FP}[A]$ we obtain the following unity 
\begin{widetext}
\begin{eqnarray}
 1 &=& \Delta_{\rm FP}\int [dU]\prod_{x}\del\left(\pd^{\mu}\am^{U^{-1}}\right) \nn \\
&=& \Delta_{\rm FP}[A^{U^{-1}}]\int [dU]\prod_{x}\del\left(\pd^{\mu}\am^{U^{-1}}\right) \nn \\
&=& \Del_{\rm FP} \int [dU]\prod_x \del\left(\pd^{\mu}V_{\mu}\right) \nn \\
&=& \int [dU][d\gamma][d\bar{\gamma}][d\beta]\exp\left[i\int d^4x\left( \beta \pd^{\mu}V_{\mu} + i\bar{\gamma}\pd^{\mu}\pdm\gamma \right) \right] \nn \\
 &\equiv&  \int [dU][d\gamma][d\bar{\gamma}][d\beta]\exp\left[i\int d^4x \left(-i\tilde{\delta}_{\rm B}\tilde{G}_{\rm GF + FP}[V_{\mu},\gamma,\bar{\gamma},\beta]\right)\right],
\label{unity}
\end{eqnarray}
\end{widetext}
where we have defined the new BRST transformation $\tilde{\del}_{\rm B}$ as 
\begin{eqnarray}
 & & \tilde{\del}_{\rm B}V_{\mu} = \pdm\gamma,~~~~\tilde{\del}_{\rm B}\gamma = 0, \nn \\
& & \tilde{\del}_{\rm B}\bar{\gamma} = i\beta,~~~~\tilde{\del}_{\rm B}\beta = 0.
\end{eqnarray}
When Eq. (\ref{unity}) is inserted, the partition function can be
rewritten as follows,
\begin{widetext}
\begin{eqnarray}
 Z_{\rm U(1)}[J]&=&\int [dU][dC][d\bar{C}][dB]\exp(iS_{\rm TQFT}[\Omega_{\mu},C,\bar{C},B]
 + iW[U;J]
 + J^{\mu}\Omega_{\mu} + J_{\rm C}C  + J_{\rm \bar{C}}\bar{C} + J_{\rm B}B),  \\
S_{\rm TQFT} &\equiv& -i\db\bdb\int d^4x \left[\frac{1}{2}\Omega_{\mu}^2 + iC\bar{C}\right],
\end{eqnarray}
where
\begin{eqnarray}
\label{2.14}
 \e^{iW[U;J]} \equiv \int [dV_{\mu}][d\gamma][d\bar{\gamma}][d\beta]\exp\left(
iS_{\rm pU(1)}[V_{\mu},\gamma,\bar{\gamma},\beta] + i\int d^4x V_{\mu}\mathcal{J}^{\mu}\right), \\
\label{2.15}
S_{\rm pU(1)}[V_{\mu},\gamma,\bar{\gamma},\beta] = \int d^4x \left(
-\fr{1}{4}F_{\mu\nu}[V]F^{\mu\nu}[V] - i \tilde{\del}_{\rm B} \tilde{G}_{\rm GF + FP}[V_{\mu},\gamma,\bar{\gamma},\beta]\right), \\
 \mathcal{J}_{\mu} \equiv J_{\mu} + i\db\bdb \Omega_{\mu}.~~~~~~~~~~~~~~~~~~~~~~~~~~~~~~~~~~~~~~~~~~~~~~~~~~~~~~ \nn
\end{eqnarray}
\end{widetext}
The action (\ref{2.15}) 
describes the perturbative 
deformation part \footnote{We must integrate the variable $V_{\mu}$ in the
partition function. Therefore, the gauge fixing term is needed in the
perturbative part in order to fix the gauge degree of freedom of
$V_{\mu}$. This part reproduces the known result in the ordinary
perturbation theory. }.
The action $S_{\rm TQFT}$ is $\db$-exact and 
describes the topological model, which contains 
the information of the confinement.

In what follows we are interested in the finite temperature
system (i.e., the system coupled to the thermal bath). 
Therefore we have to perform 
the Wick rotation of the time axis and move from the Minkowski
 formulation to the Euclidean one.

\subsection{Expectation values}

\begin{figure}
\includegraphics{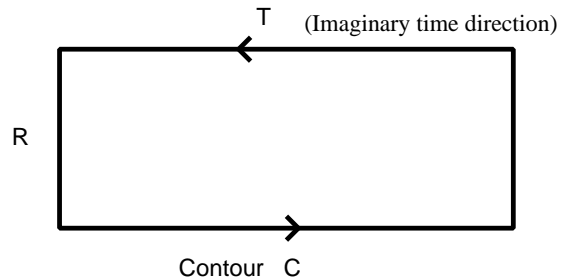}
\caption{\label{wil:fig}
The rectangular Wilson loop. This includes the
imaginary time axis in order to study the confinement.}
\end{figure}

We can define the expectation value in each sector using the action
$S_{\rm pU(1)}$ and $S_{\rm TQFT}$. The expectation value
of the Wilson loop or Polyakov loop is an important quantity to
study the confinement.  In the case of the Wilson loop $W_{C}$,
the following relation:
\begin{eqnarray}
\label{expect}
 \la W_{C}[A] \ra_{\rm U(1)} &=& \la W_{C}[\Omega] \la W_{C}[V] \ra_{\rm pU(1)} \ra_{\rm TQFT} \nn\\
& =& \la W_{C}[\Omega] \ra_{\rm TQFT}\la W_{C}[V] \ra_{\rm pU(1)}
\end{eqnarray}
is satisfied \cite{qed}. The contour $C$ is rectangular as shown
in Fig. \ref{wil:fig}. The Wilson loop expectation 
value is completely 
separated into the TQFT sector and the perturbative deformation part.
That is, we can evaluate the expectation value in the TQFT sector 
independently of the perturbative deformation part. In fact, 
we can derive the linear potential by investigating the TQFT
sector.  

\begin{figure*}
\includegraphics{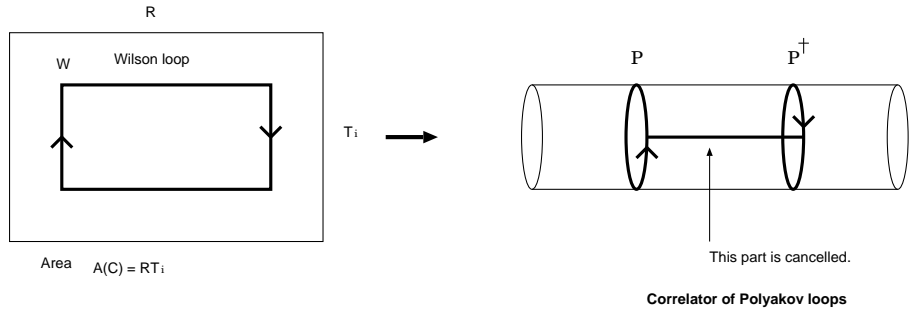}
\caption{\label{poly:fig}
The correlator of Polyakov loops. The expectation
value of a Wilson loop $W$ is equivalent to a correlator of the
Polyakov loops $P$ and $P^{\dagger}$.}
\end{figure*}
At finite temperature we must evaluate the correlator of the Polyakov
loops $P({\bf x})$. It can be evaluated in the same way as the Wilson loop, 
due to the following relation (as shown in Fig. \ref{poly:fig})
\begin{equation}
 \la P({\bf x}) P^{\dagger}(0) \ra_{\rm U(1)} = \la W_{C}\ra_{\rm U(1)}. 
\end{equation}

Furthermore, 
we can derive the Coulomb potential (at zero temperature) \cite{qed}
or Yukawa-type potential (at finite temperature) using the hard thermal
loop approximation \cite{KY2} from the perturbative deformation part.

\subsection{TQFT sector and PS dimensional reduction}

When the gauge group is  the compact $U(1)$
the TQFT sector  becomes the $O(2)$ NLSM$_2$ through the PS dimensional 
reduction \cite{PS}.  The four-dimensional TQFT sector action
\begin{equation}
 S_{\rm TQFT} = \db\bdb\int d^4 x \left[\frac{1}{2}\Omega_{\mu}^2 + iC\bar{C}\right] 
\end{equation}
can be rewritten as the $O(2)$ NLSM$_2$ on the two-dimensional space,
\begin{eqnarray*}
 S_{\rm TQFT} &=& \pi \int d^2x~ \Omega_{\mu}^2(x)\nn \\
&=& \frac{\pi}{g^2}\int d^2x \pdm U(x) \pdm U^{\dagger}(x),\nn\\
\Omega_{\mu}&\equiv&\frac{i}{g}U(x)\pdm U(x)^{\dagger}, 
\end{eqnarray*}
where we have omitted the ghost term. 
When we write the gauge group
element as $U(x) = \e^{i\varphi(x)}$, we obtain 
\begin{equation}
\label{2.22}
 S_{\rm TQFT} = \frac{\pi}{g^2}\int d^2x \pdm \varphi(x)\pdm\varphi(x).
\end{equation}
If the gauge group  $U(1)$ is not compact, 
then the TQFT sector becomes the ordinary free
scalar field theory on the two-dimensional space, 
which has no topological object. So 
 the confining phase cannot exist. 
If the $U(1)$ is compact, the theory described by the action
(\ref{2.22}) 
is the periodic boson theory. The angle variable
$\varphi(x)$ is periodic (mod $2\pi$),
 and so  $\varphi(x)$ is a compact variable. 
It is well known that 
the compactness plays an important role in the confinement \cite{Poly}.
If we consider the system at finite (zero) temperature, then the dimensionally
reduced theory lives on the cylinder (2-plane). 

We should remark here that the compactness also 
leads to monopole configurations in the original gauge theory, 
which is assumed to play an important role in the confinement. 
On the other hand, the reduced theory has vortex solutions  
due to the compactness of $U(1)$. It is quite natural that the two
configuration above 
are intimately connected, as suggested in 
Ref. \cite{qed}. Therefore we can include the monopole effect from the
reduced theory and obtain the physical quantity such
as a string tension through this sector which includes the unphysical degrees
of freedom only.  

\begin{figure}
\resizebox{.4\textwidth}{!}{\includegraphics{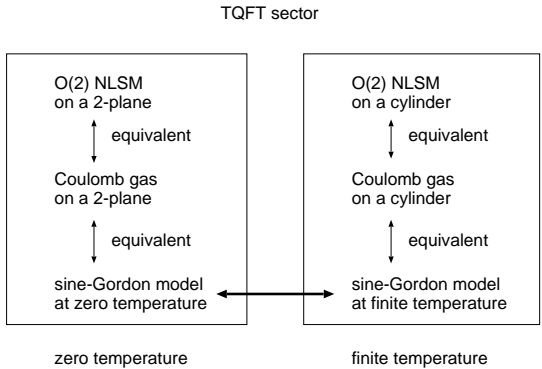}}
\caption{\label{tqft:fig}
The TQFT sector is equivalent to the two-dimensional
theory through the PS dimensional reduction. When the gauge group is
the compact $U(1)$ it becomes the $O(2)$ NLSM$_2$. The two-dimensional
space is a 2-plane (at zero temperature) or a cylinder (at finite
temperature). It is well known that it is equivalent to the several
model. }
\end{figure}
The $O(2)$ NLSM$_2$ is
equivalent to the CG system as shown in Fig. \ref{tqft:fig}.  
The partition function of the CG system is given by 
\begin{widetext}
\begin{equation}
 Z_{\rm CG} = \sum_{n=0}^{\infty}\frac{\zeta^{2n}}{(n!)^2}\prod_{i=1}^{n}\int d^2x_id^2y_i \exp\Bigg(- \frac{(2\pi)^3}{g^2}\Big[\sum_{i < j}[\Delta(x_i - x_j) + \Delta(y_i - y_j)]- \sum_{i,j}\Delta(x_i - y_j)   \Big] \Bigg),
\end{equation}
\end{widetext}
where $\zeta \equiv \exp(- S_{\rm self})$
is the chemical potential of the CG system \footnote{It is
assumed that monopoles in the original theory should be related to
vortices in the two-dimensional theory, as is discussed in Ref. \cite{qed}. 
A vortex should be be interpreted as the point that 
monopole world-line pierces the two-dimensional plane, 
which is chosen when we use the PS dimensional reduction.
We should understand that 
the chemical potential in the CG system $\zeta$ is related 
to the monopole-line self-energy.} and can be 
written in terms of the self-energy part of a vortex $S_{\rm
self}$. This quantity $\zeta$ does not depend on the physical
temperature $T$ in the original theory. 
 The $\Delta(x_i - x_j)$ expresses the Coulomb potential on the 2-plane (at
zero temperature) or on
the cylinder (at finite temperature).
The temperature of the CG system is defined by 
\begin{equation}
\label{3.8}
 T_{\rm CG} \equiv \frac{g^2}{8\pi^3}.
\end{equation}

The linear potential between the test charged
particles is induced by the effect of the vortices. 
The expectation value of the Wilson loop (Polyakov loops' correlator)
 is obtained as follows (for details, see Ref. \cite{qed}):
\begin{eqnarray}
 \la P({\bf x})P^{\dagger}(0) \ra_{\rm U(1)}& =&
 \la W_{C}[\Omega]  \ra \cong \e^{-\sigma A}, \\
 \sigma &=& \left(2\pi \frac{q}{g}\right)^2\zeta,
\label{sigma}
\end{eqnarray} 
where $A = RT_{i}$, $R=|{\bf x}|$ and $q$ is a charge of the test particles.
It is significant that the
expression of the string tension (\ref{sigma}) does not 
depend on the temperature explicitly, but on the
behavior of the CG system. 
Whether the string tension remains or vanishes is determined only by  
 the behavior of the CG system which consists of vortices. Thermal effect
changes the behavior of the topological objects. The change is
reflected to the linear potential. 
We also note that the string
tension $\sigma$ is proportional to $\zeta$. The $\zeta \rar 0$
limit implies that the self-energy of the vortex is infinity and 
we fail to include the effect of the vortex.   Therefore 
the confining phase also vanishes in the limit $\zeta\rar 0$ as is expected.

\begin{figure}
\resizebox{.4\textwidth}{!}{\includegraphics{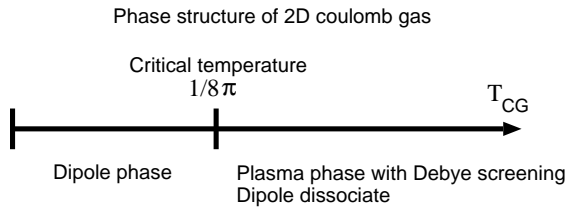}}
\caption{Two-dimensional Coulomb gas has two different phases. Over the
critical temperature $T_{\rm CG} =1/8\pi$, the system is the plasma phase with
Debye screening, and so the mass gap exists. Below the $T_{\rm CG}$ it is
the dipole phase, in which Coulomb charges form dipoles. The system
has a long-range correlation and no mass gap.}
\label{CG:fig}
\end{figure}

In our previous paper \cite{KY2} we have investigated the phase
structure of the compact $U(1)$ gauge
theory at finite temperature using the behavior of the 
CG system on the 
cylinder. The behavior of the CG system on the 2-plane is shown in
Fig. \ref{CG:fig}. The behavior of the CG system on the cylinder 
would be analogous. We expect that the system undergoes the BKT-like
phase transition.  In fact, the CG system causes the BKT-like phase
transition in the high-temperature region 
that it behaves as the one-dimensional
system. 

\section{Equivalence between SG model and CG System}

The action of the  SG model on the 2-plane (at zero temperature) or 
the cylinder (at finite temperature) is defined by  
\begin{equation}
S_{\rm SG}=
\int d^2x \left\{\frac{1}{2}\left(\pdm \phi\right)^2 - \frac{m^4}{\lambda}
\left[\cos\left(\frac{\sqrt{\lambda}}{m}\phi\right) - 1\right]\right\}.
\end{equation}
The partition function is given by 
\begin{eqnarray}
 Z_{\rm SG} &=& \int [d\phi] \exp \left(- S_{\rm SG}\right).
\end{eqnarray}
This can be rewritten as follows:
\begin{widetext}
\begin{eqnarray}
 Z_{\rm SG} &=& \e^{- (m^4/\lambda)\int d^2x}
\sum_{n=0}^{\infty}\frac{1}{n!}\left(\frac{m^4}{\lambda}\right)^n
 \int [d\phi] \e^{-\int d^2x (1/2)\left(\pdm \phi\right)^2  } \left[\int d^2x \cos\left(\frac{\sqrt{\lambda}}{m}\phi\right)\right]^n \nn \\
&=& \e^{- (m^4/\lambda)\int d^2x} \sum_{n=0}^{\infty}\frac{1}{2n!}\left(\frac{m^4}{2\lambda}\right)^{2n} \int [d\phi] \e^{-\int d^2x (1/2)\left(\pdm \phi\right)^2  } \left[\int d^2x\left( \e^{i(\sqrt{\lambda}/m)\phi} + \e^{-i(\sqrt{\lambda}/m)\phi}\right) \right]^{2n} \nn \\
&=&
\e^{ - (m^4/\lambda) \int d^2x} \sum_{n=0}^{\infty}\frac{1}{(n!)^2}\left(\frac{m^4}{2\lambda}\right)^{2n}   \int [d\phi] \e^{ - \int d^2x
 (1/2)\left(\pdm \phi\right)^2}\prod_{i=1}^{n}\int d^2x_id^2y_i \e^{i(
\sqrt{\lambda}/m)[\phi(x_i) - \phi(y_i)]}. \nn
\end{eqnarray}
Here, we note that 
\begin{eqnarray}
\left\la \e^{\int d^2x J\phi} \right\ra_{\rm SG} &\equiv& \int [d\phi]\exp
\left[\int d^2x \left(-\frac{1}{2}\left(\pdm\phi\right)^2 + J\phi\right)\right] \nn \\
&=& \exp \left[\int d^2x d^2y J(x)\Delta (x-y) J(y)\right],
\label{1.3}
\end{eqnarray}
\end{widetext}
where $\Delta(x-y)$ is the massless scalar field propagator,
\begin{equation}
 \Delta(x-y) \equiv \int \frac{d^2p}{(2\pi)^2}\frac{\e^{i p\cdot(x-y)}}{p^2}.
\label{1.4}
\end{equation}
In particular, if 
we choose the external field as 
\begin{equation}
J(x) = i \frac{\sqrt{\lambda}}{m}\sum_{i=1}^{n} q_i \del(x-x_i),
\end{equation}
then we obtain  
\begin{widetext}
\begin{equation}
\left\la \prod_{i=1}^n \e^{\int d^2x iq_i \phi(x_i)}  \right\ra_{\rm SG} = 
\left\{
\begin{array}{cc}
\exp\left[-\frac{\lambda}{2m^2}\sum_{j,k}q_jq_k\Delta(x_j - x_k)\right] & {\rm for}~\sum_{i=1}^{n}q_i = 0,  \\
0 & {\rm for}~\sum_{i=1}^{n}q_i \neq 0. 
\label{for}
\end{array}
\right.
\end{equation}
\end{widetext}
If the net charge is not zero then the correlation function vanishes because of the symmetry under the transformation 
\[
\phi(x) \longrightarrow \phi(x) + ~{\rm const}.
\]
By the use of Eq. (\ref{1.3}), we obtain the following expression of the
partition function, 
\begin{widetext}
\begin{equation}
Z_{\rm SG} \sim\sum_{n=0}^{\infty}\frac{1}{(n!)^2}\left(\frac{m^4}{2\lambda}\right)^{2n}\prod_{i}^{n}\int d^2 x_i d^2y_i \exp\Bigg(- \frac{\lambda}{m^2} 
\Big( \sum_{i < j}[\Delta (x_i - x_j) + \Delta (y_i - y_j)]
- \sum_{i,j}\Delta(x_i - y_j) \Big)\Bigg).
\end{equation}
\end{widetext}
The factor $\e^{- (m^4/\lambda)\int d^2x }$ can be ignored 
because of the normalization of the partition function.
Thus we obtain the partition function of the neutral CG system, whose
temperature is defined by
\begin{equation}
 T_{\rm CG} = m^2/\lambda . 
\end{equation}
This is equivalent to Eq. (\ref{3.8}).
The phase transition in the SG model at the critical
coupling, i.e., the Coleman transition, corresponds to the BKT phase transition in the CG system. 

The equivalence holds on the cylinder (i.e., in the finite-temperature
case). In this time, 
 Eq. (\ref{1.4}) is replaced with the propagator on the cylinder,
\begin{widetext}
\begin{eqnarray}
 \Delta(x-y) &\sim& -\frac{1}{2\pi}\sum_{n=-\infty}^{+\infty}\ln\left[
\mu\sqrt{(x_0 - y_0 - n\beta)^2 + (x_1 - y_1)^2}\right] \nn \\
&=& -\frac{1}{2\pi}\ln\left[\mu\beta\sqrt{\cosh \left(
\frac{2\pi}{\beta}(x_1 - y_1)  \right) - \cos\left(\frac{2\pi}{\beta}(x_0 - y_0)\right)  }\right].
\label{pro}
\end{eqnarray}
\end{widetext}
Here $\mu$ is the infrared cutoff and $\beta$ is the inverse of the
physical temperature $T$. We remark that the SG model has the physical 
temperature in common with the original gauge theory. It is because the
cylinder is chosen as the two-dimensional space 
when we use the PS dimensional reduction.
If we use the complex coordinates $w = x_1 + i x_0, \bar{w} = x_1 -i x_0,
w' = y_1 + i y_0$, and $\bar{w}' = y_1 - iy_0$, then Eq. (\ref{pro}) 
is rewritten as 
\begin{widetext}
\begin{eqnarray}
 \Delta(x-y) = -\frac{1}{4\pi}\ln \left| \e^{(2\pi/\beta)w} - \e^{(2\pi/\beta)w'} \right|^2 + \frac{1}{2\beta}{\rm Re} (w + w') - \frac{1}{4\pi}\ln\left( \frac{1}{2}(\mu\beta)^2\right).
\label{inf}
\end{eqnarray}
\end{widetext}
The propagator above involves the divergent term in the limit $\mu \rar$
0, but this is removed by the neutral condition 
$\sum_i q_i = 0$ [see Eq. (\ref{for})].

The parameters of the CG system in the
previous section, $g$ and $\zeta$, can be related to those of
the SG model, mass $m$ and the coupling constant $\lambda$. 
This relation is
given by 
\begin{equation}
\label{relation}
 \lambda = \frac{128\pi^{6}\zeta}{g^4},~~~~~m = 
\frac{4\pi^{3/2}\zeta^{1/2}}{g}.
\end{equation} 

Recall that the string tension $\sigma \sim \zeta$ vanishes as
$\sigma \rar 0$ and the confining phase disappears. We should note that 
$\zeta\rar 0$ means $m\rar 0$ and $\lambda \rar 0$. That is, the SG
model becomes free scalar field theory in this limit. 
This result is consistent with the disappearance of the confining phase.

\section{One-loop Effective Potential of SG Model at Finite Temperature}

It is well known that  
the SG model at zero temperature undergoes the phase transition at
certain coupling, which is called the Coleman transition \cite{Coleman}. 
The critical coupling is $\lambda/m^2 = 8\pi $, at which the quantum SG model undergoes a phase transition from a stable vacuum to an unstable 
vacuum. Moreover, the existence of a phase transition 
due to the thermal effect has been shown in Ref. \cite{f-sine}. 
This
transition would  correspond to the BKT-like transition of the 
CG system. Thus we can investigate the phase structure of the
compact $U(1)$ gauge theory at finite temperature, at least in the
region where we can investigate the SG model reliably. In particular, we
can read off from Eq. (\ref{relation}) that
the weak-coupling region of the SG model 
corresponds to the strong-coupling region of the 
gauge theory. That is, we can investigate the phase structure of the gauge
theory with the strong
coupling from the perturbative study of the SG model. This is the
advantage of our method.

\begin{figure}
\resizebox{.4\textwidth}{!}{\includegraphics{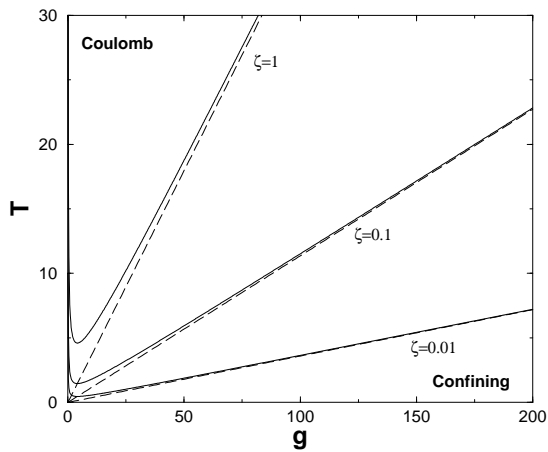}}
\caption{The phase structure of the compact $U(1)$ gauge
 theory obtained by the one-loop effective potential calculation in 
the SG model. The critical-line Eq. (\ref{cr-eq}) can be  numerically
 solved. The asymptotic line Eq. (\ref{grad}) is also drawn with the
 dashed line. 
The above graph shows that the confining phase vanishes in
 $\zeta \rar 0$ limit. However, the result in the small $g$ region is
 not valid. It is  
because the small $g$ means the large $\lambda$ and 
the one-loop approximation in the SG model is not reliable.} 
\label{plot:fig}
\end{figure}

We will discuss the one-loop effective action of the two-dimensional SG
model at finite temperature \cite{f-sine}. 
The effective potential is given by
\begin{widetext}
\begin{eqnarray}
\label{4.1}
 V_{1{\rm loop}}(\phi_c) &=& V_{\rm 0}(\phi_c) + V_{\rm FT}(\phi_c),  \\
\label{4.2}
V_{\rm 0}(\phi_c) &\equiv& \frac{m^2}{8\pi}\cos\left(\sqrt{\lambda}\phi_c/m\right)\left[1 - \ln\cos \left(\sqrt{\lambda}\phi_c/m\right)\right], \\
\label{4.3}
V_{\rm FT}(\phi_c) &\equiv& \frac{1}{\pi\beta^2}\int^{\infty}_{0}dx \ln\left[1 - \exp\left(- \sqrt{x^2 + M^2(\phi_c)\beta^2}\right)\right], \\
M^2 (\phi_c) &\equiv& m^2 \cos\left(\sqrt{\lambda}\phi_c/m\right).
\end{eqnarray}
\end{widetext}
The second equation (\ref{4.2}) is the temperature-independent part, and
the third equation (\ref{4.3}) is the temperature-dependent part which 
vanishes in the zero-temperature limit $\beta \rar \infty$ . Also, the minimum of the potential $\phi_c = 0$ is still
stable under one-loop quantum fluctuations at zero temperature.
 Taking the second derivative of Eq. (\ref{4.1}) with respect to $\phi_c$
 at $\phi_c =0$, we can evaluate a critical-line equation
 \cite{critical} as 
\begin{eqnarray}
 m^2(\beta) &=& m^2 + \frac{\pd^2 V_{\rm FT}}{\pd\phi_c^2}(\phi_c =0) \nn \\
&=& m^2 - m^2\frac{\bar{\lambda}}{2\pi}f(\bar{\beta})  = 0,
\end{eqnarray}
where $\bar{\lambda} \equiv \lambda /m^2$, $\bar{\beta} \equiv \beta m$, and $f(\bar{\beta})$ is defined by 
\begin{equation}
 f(\bar{\beta}) \equiv \int^{\infty}_0 dx \frac{1}{\sqrt{\bar{\beta}^2 + x^2} \left[\exp\left(\sqrt{\bar{\beta}^2 + x^2} \right) - 1\right]}.
\end{equation}
That is, the critical-line equation is given by 
\begin{equation}
\label{cri}
 1 - \frac{\bar{\lambda}}{2\pi}f(\bar{\beta}) = 0.
\end{equation}
Note that the parameters of the SG model can be replaced with the ones of 
the compact $U(1)$
gauge theory using the relation (\ref{relation}).  As the result, we
obtain the relation
\begin{equation}
 \bar{\beta} 
= \frac{4\pi^{3/2}\zeta^{1/2}}{gT},~~\bar{\lambda} = \frac{8\pi^{3}}{g^2}.
\end{equation}
Thus the critical-line equation can be rewritten as follows, 
\begin{equation}
\label{cr-eq}
 1 - \frac{4\pi^{2}}{g^2}f\left(\frac{4\pi^{3/2}\zeta^{1/2}}{gT}\right) = 0.
\end{equation}
The numerical solutions of this equation at various fixed values of 
$\zeta$ are shown in Fig. \ref{plot:fig}. 

In particular, we can derive the simple relation between $\lambda$ and $T$ 
in the weak-coupling and high-temperature limit of Eq. (\ref{cri}).  
The critical temperature $T_{c}$ is given by 
\begin{equation}
\label{temp}
 T_{c} = \frac{4m^3}{\lambda}.
\end{equation}
Equations.(\ref{relation}) and (\ref{temp}) lead to   
\begin{equation}
\label{grad}
 T_{c} = \frac{2 \zeta^{1/2}}{\pi^{3/2}}g.
\end{equation}
Equation (\ref{grad}) is drawn as the dashed line in Fig. \ref{plot:fig}.
Note again that the weak-coupling and high-temperature region in the SG model
corresponds to the strong-coupling and high-temperature region in the
compact $U(1)$ gauge theory. The compact gauge theory is
related to the SG model by a kind of S-duality in our scenario. 
Thus we can
reliably investigate the strong-coupling region of the gauge
theory since the
one-loop effective potential is appropriate in the weak-coupling region.

In conclusion, we have obtained that
 the critical temperature is proportional to the coupling
constant of the compact $U(1)$ gauge theory in the strong-coupling and
high-temperature region. This result is in good agreement with the
prediction in Ref. \cite{SV}. In the $\zeta \rar 0$ limit the
gradient in Eq. (\ref{grad}) goes to  zero. This fact implies 
that the confining phase disappears. 

{\it Comment on the effective potential calculation.} 
Our discussion in this section is closely analogous to Ref. \cite{HT} in which
the critical temperature has been estimated by the calculation of the 
 one-loop effective potential in the TQFT sector.
In the above discussion we have calculated the one-loop effective
potential in the SG model. If we naively calculate the effective potential in
the $O(2)$ NLSM$_2$ then we cannot obtain the phase structure  
\cite{HT,KY}. 
Note that the $O(2)$ NLSM$_2$ is equivalent to the SG
model when we consider  vortex solutions. 
If we calculate the effective potential in the $O(2)$ NLSM$_2$ we
cannot include the effect of the vortex solution. However, once we go
 from the $O(2)$ NLSM$_2$ 
to the SG model, we can include the effect of the vortex solution in
terms of the cosine-type potential. 
Therefore the phase structure that we have obtained in the SG model 
is not equivalent to  the result in the $O(2)$ NLSM$_2$. 
Moreover, the SG model is a massive
theory and the Coleman-Mermin-Wagner theorem \cite{CMW} is not an obstacle.

\begin{figure}
\includegraphics{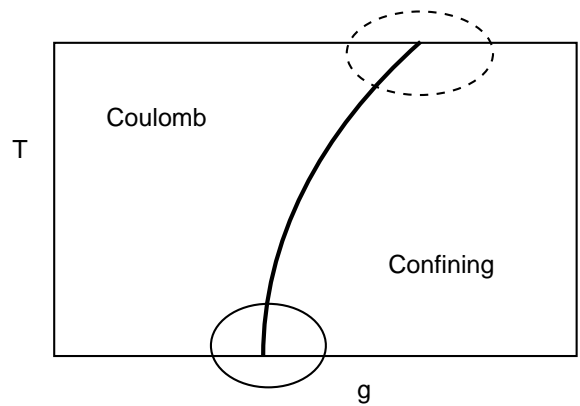}
\caption{Phase structure of the compact $U(1)$ gauge theory
 predicted in Ref. \cite{SV}.
 In the above discussion using the one-loop effective potential of 
the SG model we could study the
 region enclosed by the dashed line, in which we  can especially obtain the
 critical-line equation. Moreover, we might investigate the region enclosed
 by the solid line. It is well known that this region is neatly 
described by the
 Gaussian effective potential (GEP), which is a nonperturbative method and
 should include the physics beyond the one-loop level.}
\label{Tg:fig}
\end{figure}

\section{Conclusion and Discussion}
We have discussed the phase structure of the compact $U(1)$ gauge theory
at finite temperature by using a deformation of the topological model.
The compactness of the gauge group leads to a confining
phase. In the case of zero temperature, the phase transition of the
gauge theory 
at certain coupling 
can be described by the Coleman
transition in the SG model. 
In the finite temperature case we could investigate
the phase structure at sufficiently high-temperature and very
strong-coupling region by analyzing the one-loop effective potential of
the SG model. 
We could consider 
the enclosed region by the dashed line in Fig. \ref{Tg:fig}.
In this paper we have used the one-loop effective potential, 
but we can also use the Gaussian effective potential (GEP) \cite{st}, which is known as the
 nonperturbative method. For the SG model at zero temperature 
this is given by the following expression:
\begin{equation}
 V_{\rm GEP} = m^2 \frac{1-\bar{\lambda}/8\pi}{\bar{\lambda}}\left[1-(\cos\sqrt{\bar{\lambda}}\phi)^{1/(1-\bar{\lambda}/8\pi)}\right],
\end{equation}
where $\bar{\lambda}=\lambda/m^2$.
That is, as the coupling constant of the SG model $\bar{\lambda}\rar 8\pi$ the GEP becomes a straight line
continuously. If $\bar{\lambda}$ exceeds $\bar{\lambda}_{c}
=  8\pi$, which is the transition point of the Coleman transition, then the
GEP has the maximum and the system has no ground state.
The GEP can describe the Coleman transition
 at zero temperature \cite{I,gep-sg}, which cannot be described 
by the one-loop effective potential. Since the GEP does not depend on the
 perturbation theory it might be appropriate to investigate the phase
 structure at weak-coupling and low-temperature region in the gauge
 theory (i.e., the strong-coupling and low-temperature region in the SG 
 model) which corresponds to the enclosed region by the solid line 
in Fig. \ref{Tg:fig}. This work is  
 very interesting and will be discussed in another place \cite{KY5}. 

We should comment on the well-known results in the compact Abelian lattice
gauge theory. 
This theory  at zero temperature experiences the phase transition of
the weak first or strong second order. 
Unfortunately, our results
suggest that at high temperatures the phase transition is of the BKT
type, and do not seem to correspond to the lattice results. 
Our formalism deeply depends on the BKT phase transition, and so 
it seems difficult to predict the order of the transition obtained in the
lattice gauge theory.

It is also attractive to approach the phase structure from the viewpoint
of the massive Thirring model \cite{steer}.

\acknowledgments

The authors thank K.~Sugiyama, J.-I.~Sumi, 
T.~Tanaka, and S.~Yamaguchi 
for useful discussion and valuable comments. 
They  also would like to
acknowledge H.~Aoyama for encouragement.

\end{document}